\documentstyle [titlepage,12pt,seceqn]{article}
\def\bea {\begin{eqnarray}}
\def\eea {\end{eqnarray}}
\def\be {\begin{equation}}
\def\ee {\end{equation}}
\def\bd {\begin{displaymath}}
\def\ed {\end{displaymath}}
\begin{document}
\pagestyle{plain}
\baselineskip=0.62cm

\begin{center}
{\large 
METASTABLE DEFECTS IN THE GENERIC 2HSM 
} \footnote{To appear in the Proceedings of the XXXII-th Rencontres de 
Moriond, "Electroweak Interactions and Unified Theories", Les Arcs,
France; March 15-22, 1997.}
\end{center}

\vspace {1.2cm}

\centerline{ {\large T.N. Tomaras     \footnote{
email: tomaras@physics.uch.gr} } }
\vskip 0.4cm 

\centerline {\it {Department of Physics and Institute of Plasma Physics, 
University of Crete}} 
\centerline {\it { and Research Center of Crete }}
\centerline {\it {P.O.Box 2208, 710 03 Heraklion, Crete; Greece }} 
\vskip 1.5cm
\centerline{\large\bf abstract}

A new kind of classically stable static solitons 
called ${\it {metastable \, quasi-topological}}$ 
${\it {defects}}$ (MQTD) and a systematic method to search for 
them is presented, with examples from 
realistic particle
physics models. They are 
characterized by a topological winding number, 
which is not absolutely conserved so that the MQTD may be converted
to radiation by quantum mecanical tunneling.
The two-Higgs standard model (2HSM) supports the existence 
of classically stable membranes 
for Higgs masses consistent with present day phenomenology and
with perturbative unitarity, as well as with loop corrected 
MSSM. 
We also comment upon the possibility of metastable strings in the 
generic 2HSM.  

\newpage

\section{Introduction}
As is well known, apart from a few isolated cases in one spatial
dimension, the only known solitons are ${\it {topological}}$ solitons and
in order to  
arise either a non-trivial vacuum manifold or a non-trivial target space are 
necessary. The corresponding solitons are classified accordingly in two
classes. 

(I) The simplest paradigm of the
first is the kink solution of the model
${\cal L}=1/2 (\partial_\mu
\phi)^2 - \lambda (\phi^2-v^2)^2 /4 $ for the real scalar field
$\phi$ in 1+1 dimensions. The model
has a ${\it {non-trivial}}$ ${\it { vacuum\; manifold}}$ 
(set of degenerate minima) ${\cal M}=
\{-v, +v\}$. Field fluctuations about any of these vacua
describe particles of mass $m=\sqrt{2\lambda} v$. 
The model admits in addition 
the well-known kink solution which has finite energy and 
interpolates smoothly
between the two different vacua as one moves in space 
from $-\infty$ to $+\infty$.
Its energy is localised over a distance of order $m$. 
Furthermore, one may easily be convinced that an infinite energy barrier
separates the kink from the vacuum and guarantees its absolute stability 
\footnote{Formally, the stability of the kink is a consequence of the
conservation of the current $J_\mu = \epsilon_{\mu\nu} \partial^\nu \phi$.}.
To summarize, necessary condition for the existence of the first class
of topological solitons is ${\it {a\; non-trivial\; vacuum\; manifold}}$ 
and the solitons correspond to ${\it {non-
trivial\; maps\; of}}$ 
${\it {spatial\; infinity\; into\; the\; vacuum\; manifold}}$. 
Examples of such solitons are the topological domain walls in particle
physics and cosmology, the superfluid vortices, the Abrikosov vortices in 
superconductors and their
analogue 
cosmic strings in relativistic models. 
The magnetic monopoles
in 3+1 dimensions also belong to this class of topological solitons.

(II) The existence of ${\it {a\; non-trivial\; field\; space}}$ is the 
necessary condition for solitons of the second kind to arise, and
they correspond to  
${\it {non-trivial\; maps\; of\; space\; into\; the}}$ 
${\it {target\; space}}$.
The simplest example of a soliton of this class arises in the
sine-Gordon model 

\noindent ${\cal L}=1/2 |\partial_\mu [exp(i\Theta)]|^2 + 
\mu^2 {\cal R}e [exp(i\Theta)] = 1/2 (\partial_\mu \Theta)^2 + 
\mu^2 \cos\Theta$. 
The above model has a unique vacuum ($\Theta(x)=0$), 
a non-trivial target space given by the set $S^1$ of possible values
of the field $exp(i\Theta)$ and supports the well-known sine-Gordon
solitons, in which the angle $\Theta$ rotates by $2\pi$ as one scans
space from $-\infty$ to $+\infty$. 
The example generalizes to higher dimensions and other models with more
complicated target spaces. 
The Neel and Bloch 
domain walls in ferromagnets, the magnetic
bubbles of the ferromagnetic continuum and the Belavin-Polyakov soliton
in the $O(3)$ non-linear $\sigma$-model in 2+1 dimensions, as well as 
the skyrmions and the Hopfions in three spatial dimensions 
are examples of solitons which belong to
this second class of topological defects.

In contrast to the case of Condensed Matter systems in which a large 
variety of topological solitons have been observed and studied 
extensively, both theoretically and experimentally, no such object
has ever been observed in particle physics.   
Despite of this rather discouraging situation, thinking about such
non-perturbative aspects of particle physics has often led to 
interesting physical phenomena and new ideas about our Universe. 
The phenomenon of catalysis of proton decay in the presence of a GUT
monopole and the initial conception of the idea of inflation
which changed our view of the very early universe, are two 
examples of recent development triggered by the study of the physics
of topological solitons.

Below, I will introduce a new class of defects in theories 
with ${\it {trivial\, target
\,space\, and\, vacuum}}$ ${\it {manifold}}$ which 
according to the previous discussion
do not support the existence of any sort 
of absolutely stable 
topological soliton.
The solutions I will present are static, have finite energy
and correspond to local minima of the energy functional.
They are characterized by a topological quantum number
which contrary to the case of topological solitons 
is not absolutely conserved. 
The defects may be converted to radiation through quantum mechanical
tunnelling.

\section{ The method }
The simplest system to introduce the Metastable
Quasi-Topological Defects and to present the general method 
used in their search
is a complex scalar $\Phi$ with dynamics described by 
${\cal L}=1/2 |\partial_\mu \Phi|^2 - 1/4 \lambda (|\Phi|^2-v^2)^2 +
\mu^2 v {\cal R}e \Phi$ in 1+1 spacetime dimensions \cite{rRibbons}. 
The potential is a tilted Mexican hat with a unique minimum,
while the field space being the whole
complex plane is also trivial.  
As a result the model at hand does not possess any kind of 
topological solitons.
In the limit $\lambda \to \infty$ though, the generic 
finite energy configuration
has the form $\Phi=v\, exp{[i \Theta(x)]}$, and the target space 
is reduced dynamically 
into an effective $S^1$. The dynamics of $\Theta$ is described by 
the sine-Gordon model 
with the absolutely stable topological solitons discussed above. It is
intuitively natural to expect that once we relax the parameter $\lambda$
to finite values, the soliton will not disappear altogether, 
but instead, there
will be a critical value above which a stable solution will still exist and
below which the solution will cease to exist.   
Indeed, 
it is straightforward to verify
numerically that for $m/\mu \ge 6.1$ a classically stable soliton solution
arises, which disappears for $m/\mu < 6.1$ \footnote{The only
classically relevant parameter of the model is $m/\mu$ as can be seen
by rescaling $\Phi \to v\,\Phi$ and $x \to {(1/{\sqrt{\lambda}\,v)}} x$.}.
This result has a rather obvious interpretation: The soliton under 
discussion may be pictured by a closed elastic rubber with equilibrium
length equal to zero, tied around the central
peak of the tilted potential, without friction
and with one point fixed at its minimum as required by the finiteness
of the energy. 
If the central peak is high enough (large value of the parameter $m/\mu$)
the rubber cannot slip over the tip of the potential and the soliton is
classically stable. If instead the central peak is low the rubber
slips over and shrinks to the vacuum configuration.
Of course, even the classically stable solution can decay
to the vacuum quantum mechanically. It takes finite action for 
the rubber to follow
the classically forbidden path and slip over the barrier.
The decay rate per 
unit length is exponentially suppressed by the Euclidean action
of the corresponding bounce. 

Recipe: To search for metastable solitons of the kind described above in
topologically trivial theories one first determines a limit of the
parameters in which the target space becomes non-trivial and the reduced
model leads to absolutely stable topological solitons. These solitons
generically continue to exist even after relaxing the parameters 
away from the limit, as long as their values are kept 
within a dynamically determined range sufficiently close to it.

\section{Defects in the two-Higgs Standard Model}
The potential of the generic two-Higgs doublet ($H_1$, $H_2$) 
Standard Model is \cite{rBook} 
$$
V(H_1, H_2)=\lambda_1 \bigl(|H_1|^2-v_1^2/2 \bigr)^2+
\lambda_2 \bigl(|H_2|^2-v_2^2/2 \bigr)^2+
\lambda_3 \bigl(|H_1|^2+|H_2|^2-(v_1^2+v_2^2)/2 \bigr)^2+
$$
$$
+\lambda_4 \bigl(|H_1|^2 |H_2|^2 - (H_1^\dagger H_2)(H_2^\dagger H_1) \bigr)+
\lambda_5 \bigl(Re(H_1^\dagger H_2)-v_1 v_2 cos\xi /2 \bigr)^2+
$$
$$
+\lambda_6 \bigl(Im(H_1^\dagger H_2)-v_1 v_2 sin\xi /2 \bigr)^2
$$
with all parameters $\lambda_i$ positive or zero. It is the most general
form with CP and the discrete symmetry $H_1 \to -H_1$ broken only
softly, to avoid unacceptably large FCNC effects.

The gauge group of the model is G=SU(2)$\times$U(1) broken down to
H=U(1)$_{em}$ and
it is well-known that the first and second homotopy groups of G/H are
both trivial. In the absence of any discrete symmetry from the potential 
of the model the zeroeth homotopy group is also trivial. No topological
domain wall, string or magnetic monopole exist in this theory. 
Since, in addition the target space of the model is trivial (${\cal R}^8$)
no topological solitons of the second class arise. 

${\it {Metastable\; Membranes:}}$ 
In the limit $\lambda_1, \lambda_2, \lambda_4 \to \infty$ (in which all 
Higgses except $A^0$ become infinitely massive) the model reduces to
a 3+1 dimensional sine-Gordon model, which admits the topological 
domain walls mentioned several times above. 
According to the previous discussion, one expects the model to support
the existence of such domain walls as local minima of the energy
functional even after one relaxes the Higgs masses to finite values,
as long as they stay "large enough". The determination of the 
range of values of the Higgs
masses for classically stable 
${\it {membranes}}$ to exist is a dynamical issue which was 
studied numerically
in Reference [3]. There one may find the plot of the membrane stability
region of the parameter space. It is slightly more complicated 
but one may roughly state that as long as the
mass ratios $m_h/m_{A^0}$, $m_{H^0}/m_{A^0}$ 
and $m_{H^+}/m_{A^0}$ in the standard notation \cite{rBook}, 
are all larger than 
approximately 2.2, metastable quasi-topological membranes arise
in the 2HSM. 

Although one may envisage various other possibilities, 
a naive expectation is that such a membrane will be characterised
by a mass of order ${\cal O}(10^{10} gr/cm^2)$
and a cosmological life-time per unit area. In such a case they are
going to have desastreous consequences in our Universe, exactly like
the absolutely stable topological domain walls.
The curves of the membrane 
stability region which appear in Reference [3] 
should then be interpreted as providing upper bounds on the corresponding
Higgs masses.
Thus, for example, the set of values 
$m_A=50$GeV, $m_h=125$GeV, $m_{H^0}=160$GeV
and $m_{H^+}=200$GeV, perfectly consistent with
present day experimental bounds analyzed either in the context
of the generic 2HSM as well as with the one-loop 
corrected MSSM \cite{rKane}, \cite{rTornkvist}, 
should be excluded on
cosmological grounds. 

${\it {Metastable\; Cosmic\; Strings}}$ \cite{rStrings2},\cite{rStrings}: 
In a similar fashion one may consider the limit $\lambda_1, \lambda_2,
\lambda_5 \to \infty$ (in which all Higgses except $H^+$ 
become infinitely massive).
Correspondingly, of all the components of the Higgs 
doublets only a unit vector
field remains unfrozen, and the correspoding
effective theory is the well-known O(3) non-linear $\sigma$-model
with its well-known Belavin-Polyakov topological solitons in two
spatial dimensions.  
Like in the membrane case one may argue analytically \cite{rStrings2}, 
\cite{rStrings} and verify 
numerically \cite{rStrings3} 
that these solitons, slightly deformed, continue to
exist even as one takes the parameters away but close to their limiting
values. 

It is clear from the above discussion that MQTD are guaranteed
to arise in appropriate ranges of the Higgs and gauge particle 
masses in any model with 
an extended Higgs sector. Consequently, in the case of unwanted walls
it is not enough to check that the model of interest does not
possess a spontaneously broken discrete symmetry. One has to 
verify in addition that no metastable membranes  
exist with cosmological lifetimes.
Similarly it may not be necessary to introduce extra spontaneously 
broken symmetries or carefully chosen target spaces 
in order to produce domain walls, cosmic strings 
or localized solitons. They may arise dynamically as metastable defects of
the sort discussed here \cite{rTrehagirevons}. 

{\bf Acknowledgments}

I would like to thank P. Fayet, G. Kane and M. Shaposhnikov for several
helpful discussions during the meeting 
and the organizers of the XXXII Rencontres de Moriond
for their kind hospitality.
This research was supported in part by the EU grants 
CHRX-CT94-0621 and CHRX-CT93-0340, and by 
the Greek General Secretariat
of Research and Technology grant No 95$\Pi ENE\Delta$ 1759.


\vfill

\end{document}